\documentstyle[epsf]{mn}

\title[Nested Grid Code for Cosmological Simulations]
{A Nested Grid Particle-Mesh Code for High Resolution Simulations of
Gravitational Instability in Cosmology}

\author[R.J. Splinter]
       {Randall J. Splinter\thanks{Present Address: Center for Computational 
Sciences, 325 McVey Hall, University of Kentucky, Lexington, KY 40506-0045 USA, 
randal@ccs.uky.edu} \\ 
Department of Physics and Astronomy,The University of Kansas, 
Lawrence, Kansas 66045 USA} 
\date{Received 1995 March 15; in original form 1996 January 29}

\begin{document}
\maketitle

\begin{abstract}

I  describe a nested-grid  particle-mesh (NGPM) code designed to study
gravitational instability in three-dimensions.  The code is based upon
a standard PM code. Within the parent grid I am able to define smaller
sub-grids allowing us to  substantially extend the dynamical  range in
mass and length.  I treat the fields  on the parent grid as background
fields and utilize a one-way interactive meshing.  Waves on the coarse
parent grid are allowed to enter and  exit the subgrid, but waves from
the  subgrid are precluded  from effecting the  dynamics of the parent
grid.  On  the parent grid the potential  is computed using a standard
multiple Fourier transform technique.  On the  subgrid I use a Fourier
transform technique  to     compute the subgrid  potential     at high
resolution.   I  impose quasi-isolated    boundary conditions on   the
subgrid  using the  standard  method for generating  isolated boundary
conditions,  but rather than using the   isolated Green function I use
the Ewald method to  compute  a Green function   on the subgrid  which
possesses  the  full periodicity  of  the  parent  grid. I  present  a
detailed discussion of my methodology and a series of code tests.
\end{abstract}


\section{Introduction}

Over the past decade N-body techniques have become the dominant method
for studying the clustering of mass on large  scales (see Efstathiou {
et al.} 1985, or Hockney  \& Eastwood 1981).  Direct particle-particle
(PP)  codes were the first  n-body codes which   were widely used, and
they reached a remarkable state of development as discussed by Aarseth
(1985).  These codes   tend to  be very   time consuming since    each
particle interacts  with every other  particle directly  via a $1/r^2$
force.  This technique allows for very high  spatial resolution but at
the expense of CPU time.  As the particles clump the  the small   scale 
dynamics dominate and the CPU time required
jumps dramatically.   Treecodes
are dramatically faster since they compute direct PP interactions only
locally and  the far field computations  are  performed using nodes of
particles.  These codes  do not require a mesh  and hence do not  have
the spatial resolution problems associated  with the introduction of a
grid.  For  both of these  codes the small-scale spatial resolution is
set by the finite softening  length which is  introduced to avoid  the
formation  of  tight  binary  pairs  which  would force a  significant
decrease in   the time step  and  lead  to  dramatic CPU requirements.
Particle-Mesh codes have circumvented a  number  of those problems  at
the expense   of  the  finite spatial  resolution  introduced  by  the
existence of the  grid system.  The forces  are interpolated  from the
grid and  hence they are  truncated on  small scales.  This  helps  to
eliminate the two-body effects present in PP  codes. Thus PM codes are
well suited to the study of  Vlasov-like systems where it is essential
to suppress two-body effects, such as the dark matter density field in
cosmological  simulations.   In an  attempt   to increase the  spatial
resolution of PM codes, Particle-Particle-Particle-Mesh  ($\rm{P^3M}$)
codes were  developed. $\rm{P^3M}$ codes  are a hybrid class which use
PM   methods  to compute   large-scale  forces   and  PP  methods  for
small-scale interactions.  This   modification to  PM  codes partially
helps to increase resolution.  On the other hand  they will suffer the
same dramatic slow down which occurs with PP codes as clumping becomes
significant.    Furthermore,  as   emphasized    by   Sellwood  (1987)
$\rm{P^3M}$ codes may introduce two-body  effects on small-scales thus
making  them unsuitable for use in  modeling  Vlasov-like systems. 
Pen (1994) has constructed a code which he
refers  to  as a ``Linear  Moving  Adaptive  PM Code''.  This  code is
adaptive in the traditional    sense, where the mesh  spacing   varies
according to  some  local quantity, in  his  case the  density.  As  a
consequence  of Adaptive Mesh   Refinement (AMR)  Pen  finds that  the
particles in his  code feel a self-force,  this could be a problem for
some types of  initial conditions. Katz  \& White (1993)  introduced a
``multi-mass''    technique   which   is   based   upon   the   hybrid
N-body/hydrodynamics code TREESPH, which was developed by Hernquist \&
Katz (1989).   Katz \& White (1993)  used their ``multi-mass'' code to
examine the  properties of simulated galaxy  clusters.  In this method
they use a series of  nested lattices, with  the particle mass growing
smaller as  one moves to more  finely spaced lattices.  Thus, they are
able to obtain   simultaneously  high force  accuracy  and  high  mass
resolution.  This method is similar to the  nested-grid methods that I
will discuss next. In a similar  spirit to $\rm{P^3M}$ codes where one
increases  the  force resolution without  concern  for increasing mass
resolution  Suisalu \& Saar  (1995) and independently Jessop, {et al.}
(1994) have developed multi-grid  based adaptive particle codes.  Both
codes  are capable  of adaptively  modifying   the underlying  grid to
accommodate  the varying  distribution of particles.   Suisalu \& Saar
(1995) use a full multi-grid method for the  potential solver which is
able to  refine on  regions of  high   density increasing their  force
resolution.  Jessop, {et al.}  (1994)  use a relaxation Poisson solver
and  Neumann boundary  conditions  interpolated from the parent  grid,
coupled  with a  methodology   for adaptively  creating sub-grids   in
regions where increased force resolution would be of benefit.

Nested mesh codes have been  available in the atmospheric sciences for
a  number  of  years,  and  Koch \&  McQueen    (1987) provide a  nice
introduction.   In these fields  it  was realized  that nesting a fine
mesh  within a coarse  mesh can  be a very   economical way to achieve
higher resolution  without increasing dramatically the required memory
and   CPU  resources that   an   equivalent larger  grid  would  need.
Nested mesh techniques  are one way to extend  resolution
in simulations of  large-scale structure. Peebles  (1980) first argued
that  long wavelength Fourier   modes  can   couple to much    shorter
wavelength modes increasing the power  on small scales. More  recently
Jain \& Bertschinger  (1993) have come to  the same conclusion.   As a
consequence attempts to increase  resolution by reducing the box  size
appear to be doomed from the start. Furthermore, moving from simple PM
codes to $\rm{{P^3M}}$  codes on a  similar  sized grid  appears to be
capable of increasing the resolution by only a small factor, roughly a
factor $O(10^3)$  shy of the  required resolution for full simulations
capable of addressing both the details of galaxy formation while still
being able to follow the large wavelength modes.

Over  the last several  years a number  of nested grid codes have been
developed for use in  cosmology and/or astrophysics.   Chan {  et al.}
(1986) developed a   nested  mesh code   for use   in studying  galaxy
collisions.  In the code of Chan, { et al.}  the sub-grid potential is
computed using a standard finite difference  approach and the boundary
conditions for   the sub-grid are interpolated  from  the known coarse
grid potential.   Villumsen (1987) constructed a Hierarchical Particle
Mesh (HPM) code which is similar to the code I will describe here. 
Recently, Anninos,
{et al.} (1993) have reported a nested mesh code for use in cosmology
which not  only evolves the collisionless  dark matter  on a fine grid
but performs  a hydrodynamic  calculation on  the   fine grid as  well
allowing them to follow both  the dark matter  and the baryonic matter
component.    The code  developed by  Anninos   { et al.} is  
similar to that developed by Chan { et al.} in the sense that boundary
conditions for  the  sub-grid region  are obtained  directly  from the
parent grid by interpolating potential  values on parent mesh cells to
the sub-grid.   
In  a related approach   to the nested   grid schemes discussed  above
Couchman (1991) has modified the standard ($ {\rm  P^3 M} $) algorithm
by   introducing  refined meshes  in  regions  of  high  density.  The
traditional complaint against  such codes has  been that as clustering
evolves they tend to slow down because there  are an increasing number
of particles within one cell of one another.  This  causes the PP step
to become the dominant  portion  of the  code and significantly  slows
down the  code as  a  whole.  To circumvent  this,  in regions of high
density Couchman introduces refined meshes  to guarantee that there is
never a very large number   of particles within   the same grid  cell.
This prevents the  PP portion of the  code from dominating the overall
runtime.

This   implementation of a    nested-grid algorithm  has a  number  of
advantages over the methods previously mentioned. Unlike codes such as
${\rm P^3M}$,  and  tree  codes  I am able  to  get   very high  force
resolution  but without the  slowdown   associated in those codes  for
highly clustered distributions.  Furthermore,  I  am able to  not only
increase  the force resolution, but   also the mass resolution on  the
sub-grid region. This is something  which only other nested-grid codes
are  at present  capable of  doing. In  comparison  to the nested-grid
codes,  I  feel that  my mass advection  scheme  should make my method
easier to   generalize to a  more  adaptive scheme. I also   enforce a
Courant-Friedrichs-Levy  (CFL)  condition on the   sub-grid time step,
guaranteeing that  the time integrator does  not go unstable.  This is
contrary to the  Villumsen (1988) version  of a nested-grid code where
no CFL condition is enforced on the sub-grid particles.  

This paper is organized in the following  way: section \S II discusses
in detail the  algorithm that   I have  implemented, Section \S   III
presents the results of a number of tests  of the code, and finally in
section \S IV I present my conclusions.

\section{General Aspects of the Code}

The general  philosophy of the   code is quite  simple.   I am free to
define a sub-grid    anywhere  within the  parent   grid.   As  parent
particles move into the  sub-grid region I view  them as consisting of
$n_{sg}^3$ sub-grid  particles which initially are  on a cubic lattice
the size of one  parent grid cell.   As the parent particle enters the
sub-grid the sub-grid  particles are free to  move on a sub-grid  mesh
which is typically much finer than the parent mesh.  This allows us to
gain   significantly  better length  resolution,  albeit   in  a small
localized  region.  Furthermore, since  the  typical sub-grid particle
mass   is $m_{sg} =  m_{pg}/n_{sg}^3$,  where  $m_{pg}$ is  the parent
particle mass I am able to gain additional mass resolution beyond that
possible with the original PM code.

Up to this  point the discussion has focused  primarily  on methods to
increase force resolution. As emphasized by  Melott \& Shandarin 1989,
and  Melott 1990,  increasing the  force  resolution is not  enough in
cosmological simulations.  The authors argue that increasing the force
resolution without  a corresponding  increase  in the  mass resolution
will lead to  the growth of  spurious perturbations due to  shot noise
from the particles.   As a consequence any  method for increasing  the
force resolution without increasing  the mass resolution will tend  to
lose phase information, and thus is not getting the physics correct in
the model.   Unfortunately,  this error   will   not show up  in   the
autocorrelation under most circumstances (Melott 1990).  
Therefore, results may very
well be incorrect, but as long as the autocorrelation is being used to
study the  system the errors may  not be apparent. One great advantage
of  a nested-grid scheme over other  methods which try to increase the
force resolution without increasing their mass resolution such as tree
codes or ${\rm   P}^3{\rm M}$, is  that  a nested-grid  code will also
increase the mass resolution since  the sub-grid particles will always
have a smaller mass than the parent grid particles. A nested-grid code
will   be capable    of     increasing force    and   mass  resolution
simultaneously, thus avoiding the introduction of any phase errors due
to the growth of spurious perturbations in the simulation.

\subsection{Grid Layout and Parent Grid/Sub-grid Interfaces}

I have adopted  an adjacent mesh  structure,  as discussed  by Koch \&
McQueen   (1987),   for    the   sub-grid/parent   grid   hierarchy. A
two-dimensional  example is show in Fig.  1.  I am  free to define the
size  of  the  sub-grid in   terms  of  parent  cell  sizes, once  the
coordinates of  the lower left hand  corner of the  sub-grid have been
stated.   One can  use an arbitrary  number  of parent  grid cells and
sub-grid cells, with a uniform spacing on both grids. In addition, the
sub-grid region is fixed and is not  allowed to move as the simulation
progresses.

\begin{figure}[t] 
     \epsfxsize = 3.0truein
     \hskip 1.5truein
     \epsfbox{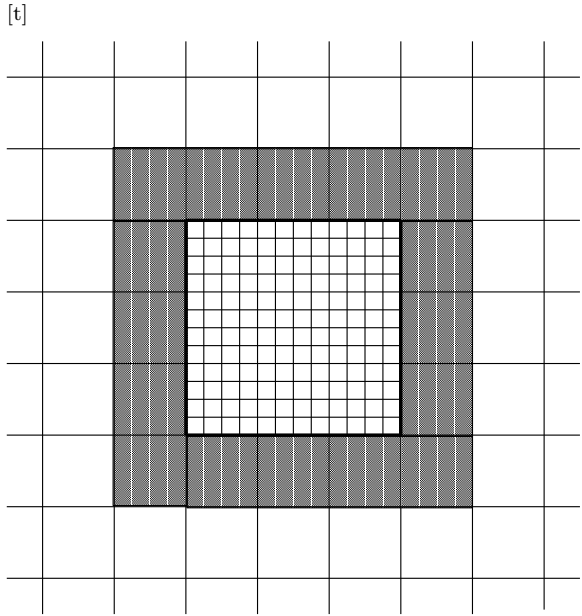}
     \caption{An example of the adjacent mesh structure that
              the nested mesh code utilizes.}
\end{figure}

In my code a  one-way interface is utilized   so that the  parent grid
particles   can influence  the   sub-grid particles,  but the sub-grid
distribution cannot back-react with the  parent particles.  Anninos  {
et al.} (1993)   found  that significant  noise was   generated when a
two-way interface  was used. As discussed   by Koch \&  McQueen (1987)
this may not  be  unreasonable to expect  since the   higher frequency
waves present  on  the   sub-grid may  generate  false  waves   at the
interface, or since the high  frequency waves cannot be represented on
the much coarser parent grid   they may lead to significant   aliasing
effects.

As emphasized  by Koch \& McQueen  (1987) and Anninos {  et al.} it is
necessary to include a buffer  zone around the  sub-grid to smooth any
density or force transients  which may be  introduced as  the sub-grid
particles enter the sub-grid region.  I  have found that a buffer zone
the width of two parent cells  is sufficient to minimize any transient
effects from mass movement into or out  of the sub-grid, but the width
of the buffer zone is an  input quantity and  it can be varied if need
be. In addition, the buffer  zone is divided in  half creating a inner
and outer  buffer  zone. The  sub-grid particles  are moved using  the
parent grid potential until they enter the  inner buffer zone. At that
point   the sub-grid  particles are   moved using  the high resolution
potential of the sub-grid region. The use of an inner and outer buffer
zone helps the  sub-grid particles to  slowly relax to  the additional
power  which  the sub-grid potential  possesses,  and therefore by the
time the sub-grid particles have reached the actual sub-grid they have
relaxed and are  moving under the  influence of the sub-grid potential
only. This helps to  eliminate any edge effects  which might come into
play  should the sub-grid particles not  be allowed  to relax prior to
their entering the sub-grid region.

Furthermore, I have found the need to add  a layer of 3 sub-grid cells
in the  buffer zone. Thus, increasing  the mesh upon which the density
assignment is performed by 6 grid cells in each  direction so that the
entire sub-grid is  increased by  two parent  cells in all  directions
plus  an additional 3 sub-grid  cells in all directions.  The addition
of these extra sub-grid cells helps to guarantee that the advection of
sub-grid particles  into the sub-grid  will not introduce any unwanted
density fluctuations at the sub-grid/parent grid boundary.  Details on
how the  forces  are  computed on  particles  as they   move  into the
sub-grid region are given below.

\subsection{Force Calculation}

The forces on the parent particles  are obtained by a simple two-point
differencing of the potential on the parent grid. To obtain the forces
on the  sub-grid I break the   calculation up into  two  stages (for a
single nested mesh).  First, I compute  the density on the parent grid
and set the density over the sub-grid region equal to the mean density
on the  sub-grid.  By setting the  density equal to  the mean over the
sub-grid  region I erase all of  the fluctuations on the sub-grid, but
the total mass present on the sub-grid is still present.  This density
field is then   Fourier transformed and  convolved  with  the standard
seven-point approximation to the Green function in $k$ space, i.e.

\begin{equation}
G(\vec{k}) = { {\theta} \over {6 - 2 \cos({{\pi k_x} \over L})
                 - 2 \cos({{\pi k_y} \over L})
                 - 2 \cos({{\pi k_z} \over L})} }
\end{equation}

\noindent  where $\theta$  is a dimensionless parameter
defined to be  $\theta \equiv 4 \pi G  \bar{\rho} \delta t^2$, and $G$
is Newton's constant, $\bar{\rho}$ is the global mean density, $\delta
t$ is the magnitude of  the time step.   As discussed in Potter (1973)
$\theta < 0.5$ for stability.  I have chosen $\theta$ to be 0.49 \ .

When this potential  is inverse Fourier  transformed it represents the
contribution to  the  subgrid potential  from  those  parent particles
which have not entered the subgrid.  I  then difference this potential
and interpolate using a CIC scheme to compute the background forces on
the subgrid  particles.  This leaves us with  the final set of forces,
the forces on the  subgrid particles due  to themselves.  Here I use a
transform  technique similar  to Villumsen   (1989),  but rather  than
imposing isolated boundary  conditions on the  subgrid potential I use
``quasi-isolated'' boundary conditions.

To test the relative error that would be made by using purely isolated
boundary  conditions as opposed   to the  ``quasi-isolated'' boundary
conditions  which  I use  in this code,   I compute the relative error
between the  Ewald  generated Green  function and the  isolated  Green
function.  The horizontal axis is the quantity $r/L$ where $r$ 
is the radius and $L$  the box size. The vertical axis is the relative
difference between the Ewald Green function and the isolated Green function.
Therefore for a sub-grid size which is $1/4$ of the parent grid the
the largest radius is $0.25 \sqrt{3}$ and the 
error associated with using the isolated Green function rather than
the Ewald Green function is roughly $40\%$. This plot is shown in Fig 2.

\begin{figure}[t] 
     \epsfxsize = 3.0truein
     \hskip 1.5truein
     \epsfbox{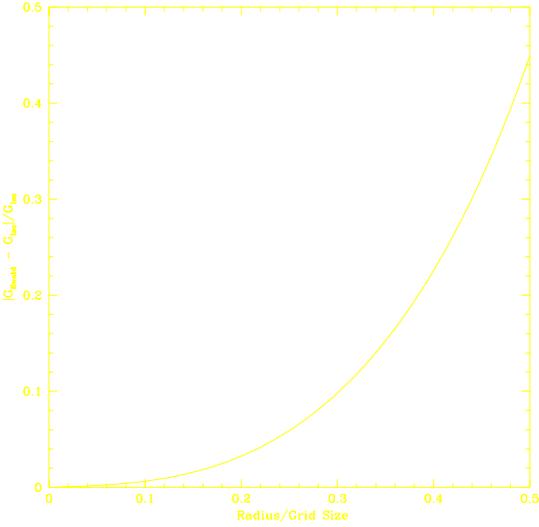}
     \caption{A plot of the relative error between the Ewald generated
     Green function and the isolated Green function.} 
\end{figure}

Eastwood  \& Browning (1979) and  Hockney \& Eastwood (1981) provide a
method for imposing isolated  boundary  conditions  on the  a  cubical
mesh.  I impose isolated boundary conditions on the sub-grid using the
method discussed  by the above authors. This  consists of doubling the
density field, padding the extra regions with zeros, and convolving it
with a Green function  which has an imposed   periodicity of the  main
grid length.  The Green function on the  subgrid is computed using the
Ewald method (1921) which possesses the full periodicity of the parent
grid.  In this  manner I take into account  the images of the sub-grid
particles   which  would  not   exist  if I  used  the  Green function
appropriate   for  an isolated   system.  This  is     what I mean  by
``quasi-isolated''  boundary   conditions.  The   Green   function  is
computed  once at  the beginning of  the  simulation, transformed  and
saved.   The transformed density is  convolved with the Green function
obtained from the Ewald method and Fourier transformed back to give us
the high  resolution subgrid  potential.   This potential can  then be
differenced  and interpolated to obtain  the  sub-grid contribution to
the forces on the subgrid particles.

Using the derivation of Ziman (1972) or Sangster \& Dixon (1976) the
Green function can be computed using the Ewald method. It takes the 
following form,

\begin{eqnarray}
G(\vec{x}, \vec{n}L) & = & { {2 \alpha} \over { \sqrt{\pi}}}  \nonumber\\ 
                     & + &{1 \over {\pi L}} \sum_{\vec{h}} \cos( {{2 \pi}
                         \over L} \vec{h} \cdot \vec{x}) 
                         { {e^{-\pi h^2/ L \alpha}} \over h^2}  \\
                     & + &\sum_{\vec{n}} {1 \over {| \vec{x} - \vec{n}L|}}
                         {\rm erfcc}( \alpha | \vec{x} - \vec{n}L| ) \nonumber 
\end{eqnarray}

\noindent  where $\vec{k} = 2 \pi  \vec{h} /  L$,  and $\vec{h}$ is an
integer triplet, and $\alpha$ is an  arbitrary parameter introduced by
the Ewald method and represents the length scale at which the real and
$k$   space expansions become important.    Hernquist { et al.} (1991)
found adequate convergence when $\alpha = 2 /  L$, $|\vec{h}|^2 < 10$,
and $|\vec{x} - \vec{n}L| < 3.6L$. I adopt their values in this code.

\subsection{Time Integration}

The  time  integration  on both the   subgrid  and the parent  grid is
performed using   a standard  leap-frog  technique. The  advantages of
using the leap-frog method are discussed in Hockney \& Eastwood (1981)
and Potter  (1973). In  order to   ensure  stability of  the leap-frog
integration  I use separate time   steps for the  parent  grid and the
subgrid. I discuss  this choice in more detail  below.   On the parent
grid the  time step is constrained  first by a Courant-Friedrichs-Lewy
(CFL) condition,

\begin{equation}
\delta t_{cfl} = 0.25 \times { {\delta x} \over {max | \vec{v} |} },
\end{equation}

\noindent then secondly by a free fall time constraint

\begin{equation}
\delta t_{ff} = { 0.49 \over { \sqrt{4 \pi G a^2 \rho_{max}}}}.
\end{equation}

\noindent The parent grid time step is  then defined to be the minimum
of the above two estimates $\delta t_{pg} = {\rm min} (\delta t_{cfl},
\delta t_{ff})$.  On the subgrid the  same two quantities are computed
and  the minimum of the two  chosen, $\delta t_{sg}$.   Then I compute
their ratio and truncate it.  This is the number of subgrid steps,

\begin{equation}
N_{steps} = {\rm int} (\delta t_{pg}/ \delta t_{sg}), 
\end{equation}

\noindent I am  allowed to take before I must  take another parent 
step to update the  parent  grid potential.   Then the actual  subgrid
time     step size is    chosen  to   be    $\delta  t_{sg}  =  \delta
t_{pg}/N_{steps}$.

The use of asynchronous  time steps for  the sub-grid particles is  an
improvement over using the same time step for both grids.  The reasons
for that  are straightforward.   I  have chosen to  use  the leap-frog
integration scheme to perform the  time integrations, therefore I must
enforce a CFL condition on both the sub-grid and the parent grid. This
is to guarantee  the stability of  the time integration. At this point
several approaches  could be used. The first  would be to  compute the
most restrictive time step on  both the sub-grid  and the parent  grid
from the CFL conditions on each, and evolve both meshes using the same
time step.  This would lead to large overhead  since the sub-grid time
step will nearly always be  much smaller than   the parent time  step,
thus forcing me to evolve the parent grid particles for many more time
steps than is  needed for an  accurate solution. Secondly, I could use
asynchronous   time steps, updating   the parent  particles only  when
necessary.  I can justify the use of an  asynchronous time step by the
following   two  arguments. First  the gravitational  infall timescale
should be an estimate of the dynamical timescale for the system. Since
the gravitational  infall timescale is  proportional to the inverse of
the square   root of the density,  and  since I  have much higher mass
resolution on the sub-grid I expect much smaller infall timescales for
the sub-grid, in other words the sub-grid particles should be evolving
on a smaller timescale. Thus to accurately  follow their motion I need
a smaller time step.   
A second justification for the use  of asynchronous time steps is also
possible  using several recent  approximation methods.  Based upon the
work of Brainerd {et a.} (1993) and Bagla \& Padmanabhan (1994) 
 on the `Frozen Potential' formalism we
know  that the potential   does not vary  too  much over the course of
evolving the  model.   Furthermore, Melott  {et al.}  (1995) argue  in
their `Stepwise    Smoothed Potential' formalism    that gravitational
clustering  is equivalent    to smoothing the   initial  potential  on
increasing scales. As a consequence objects which are distant and have
been smoothed will tend to have a small effect on the evolution of the
sub-grid distribution of  particles. Thus, I  should be able to evolve
the sub-grid distribution separately from the parent grid safely.

\subsection{Mass Advection}

Crucial to any nested mesh code is how one deals with the advection of
mass into the subgrid region.  The technique that I have chosen to use
is significantly  different than either  Villumsen (1987) or Anninos {
et al.}    (1993) who  perform   a background  run with    only parent
particles, tag  the parent particles  which actually enter the subgrid
then  during a subsequent run they   evolve the distribution of parent
particles and the  subgrid   particles whose parents  will  eventually
enter  the  subgrid.   Prior to  the   subgrid particles entering  the
subgrid they are moved using the background potential used to move the
parent particles.  Then upon  entering  the subgrid the  finer subgrid
potential with  its higher  Fourier  components is  used  to  move the
subgrid particles. A possible disadvantage  of this method is that one
must evolve  the full distribution of   sub-grid particles, both those
particles  on the  sub-grid and those  that have  not yet entered  the
sub-grid. I believe that my  scheme is significantly better, because I
avoid the  overhead entailed by  evolving those sub-grid particles not
on the  sub-grid is nonexistent, and since  the  sub-grid particles do
not  exist until they  are laid down on  the sub-grid region my method
will make  it significantly easier  to  generalize my code  to a fully
adaptive setting where the code defines local  sub-grid regions on its
own  for refinement and subsequently  lays down the sub-grid particles
without the  need for running  the code multiple  times  refining in a
local region each runtime.

The method I  use for laying down the  sub-grid particles  is somewhat
similar to  how particles are  smeared out in  a anisotropic manner in
adaptive or tensor smooth particle hydrodynamics algorithms ( Martel {
et al.}  1994,  or Benz  \& Davies  1993).    As the parent  particles
approach the  buffer  zone I   consider their  volume  as having  been
smeared out in space, with the cloud for a parent particle being given
by half the distance to the nearest parent particle in that particular
direction. As a consequence as  far as the sub-grid particle advection
algorithm is concerned  the parent particles no  longer have a  volume
equal to a parent grid cell cubed. In essence I am distorting the volume
of the sub-grid particle cloud in phase space to be consistent with the
phase space distortions of the parent particles. 
Once the characteristic size of the
parent particle is known the coordinates of the sub-grid particles can
be computed simply, assuming that they are laid down uniformly in each
direction. 

The $x$, $y$, and $z$ components of  the velocity are then assigned by
performing  a  multivariate  interpolation between the   nearest $3^3$
parent particles in phase space  in each direction.  The interpolation
algorithm I use  was developed by Renko  (1988). It is a  multivariate
interpolation scheme which is  capable of performing the interpolation
on scattered data points.  Then the  subgrid particles are moved using
only the parent potential until they reach the edge of the buffer zone
discussed  above     at  which point    they    are moved   using  the
high-resolution  potential generated on the  sub-grid.  This should be
adequate for   most  problems of  interest,  and the  following  tests
demonstrate that little significant error is introduced in the evolved
sub-grid particle  distribution.  Furthermore, I believe  the strength
of  my technique is   that it should  generalize to  a fully  adaptive
nested grid  code somewhat easier  than the Villumsen (1988) algorithm
which requires they be laid down before  evolution of the distribution
begins.

\section{Initial Conditions}

It has become standard place  in cosmological simulations to  generate
initial conditions by using  the Zel'dovich (1971) approximation.  The
Zel'dovich approximation was  first  used  to by Klypin  \&  Shandarin
(1983),  and  Efstathiou   {et al.}  (1985)  to  generate  the initial
conditions for the particles in  a cosmological simulation.   Randomly
generated  initial  particle  positions  generate  shot-noise in   the
initial density fields on all scales.  In  models close to equilibrium
at the beginning, the initial  behavior is dominated by the collective
response to  this initial noise,  thus  masking the  true modes of the
system. Traditionally  this noise has been reduced  by turning to {\it
quiet starts} to generate initial conditions.  Byers and Grewel (1970)
showed that arranging the particles uniformly at the outset suppressed
the  amplitude of  spurious  density fluctuations  due to  shot-noise.
Further, provided there are at least as  many particle planes as there
are grid spaces  in each direction  Melott (1983) and Efstathiou  { et
al.}  (1985)  showed that  a regular  grid  spacing in each coordinate
direction will  give a quiet start to  a cosmological  simulation on a
Cartesian grid.

The initial  conditions for the parent  particles are generated in the
standard way: the particles  are laid down on a  uniform mesh and then
perturbed away from the uniform mesh through the use of the Zel'dovich
approximation (1970) which requires the initial power spectrum for the
density perturbations.  The  use of  a uniform  mesh  ensures that the
shot-noise fluctuations    in the   density will   be  minimized.  The
resulting density perturbations are  consistent with the initial power
spectrum with random phases.

Villumsen (1989) has discussed a method for generating power on 
scales not resolvable by the parent grid. I have begun writing such
a code and will present the tests of the method in an upcoming 
paper, and focus only on the evolution code in this paper.

\section{Code Tests}

In this section I wish to present a number of tests of the code. 

\subsection{Force/Potential Accuracy}

The first non-trivial test, compares the accuracy of the forces on the
parent grid and on the sub-grid. One parent  particle is laid down the
parent grid with  the $4^3 =  64$ sub-grid particles  laid down at the
same grid location on the sub-grid. The sub-grid  is chosen to be five
parent cells on a side with $25^3$ mesh cells  for a refinement factor
of five, and a sub-grid mesh size of 0.2.  The peculiar choice for the
number of sub-grid cells is made so that the 64 sub-grid particles lie
in one sub-grid cell  only and are not  smeared over the eight nearest
neighbors when the density assignment is performed. The density fields
are computed  on each grid, and  the corresponding  potential field is
computed for each grid.  Then the forces are computed at 2000 randomly
chosen positions on each grid. The results are plotted in Fig.  3, the
small dots are the  forces from the parent  grid, the open circles are
the forces  determined for the refinement  given above, and the filled
circles are the forces for a refinement factor  of 15.  I note several
things concerning  this result. First,  both force fields have roughly
the  resolution  expected; the parent  grid forces   go too  soft at a
radius of roughly   1, where I  would  have expected them to,  and the
sub-grid forces go too soft at roughly  a radius of 0.2, again exactly
where I would have expected   them to.  Secondly, the sub-grid  forces
appear  to be much more   isotropic than the  parent  grid forces,  as
measured by the   typical  scatter of  the  points at  a  given radii.
Finally  the   force  resolution  on  the  sub-grid  appears to  scale
correctly as one increases the resolution of the code.

\begin{figure}[t] 
     \epsfxsize = 3.0truein
     \hskip 1.5truein
     \epsfbox{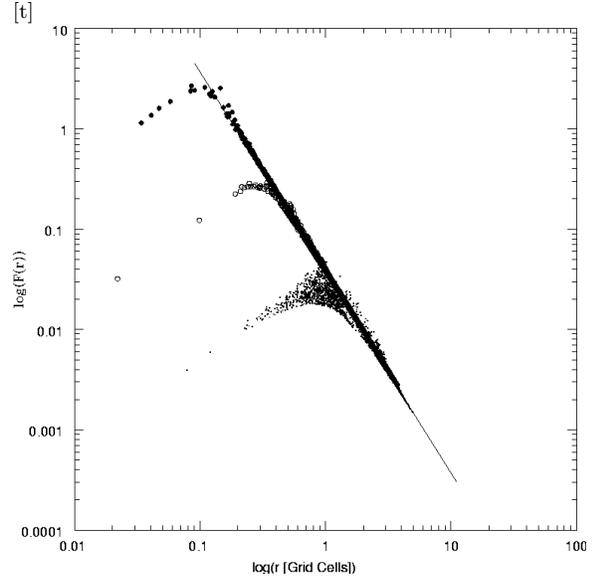}
     \caption{A plot of the sub-grid and parent grid force 
              resulting from a single particle. The parent grid 
              forces are  given by the dots, the open circles 
              give the sub-grid forces for a refinement of 5, the solid
              circles give the sub-grid forces for a refinement of 15, 
              and the solid line gives the exact 
              analytical solution from the Ewald method.}
\end{figure}

\subsection{Plane Wave Collapse - The Zel'dovich Approximation}

The simplest  and most straightforward  test is a one dimensional test
which is useful since it possesses an analytic solution.  The analytic
solution was completed by Zel'dovich in 1970, and has come to be known
as  the Zel'dovich approximation. The test has been performed with 
the perturbations aligned along the $x$ axis similar to 
Efstathiou, {et al.} (1985) for a strict PM and ${\rm P^3M}$ code. 
To make the test more challenging I run the test with the perturbation
moving diagonally from lower corner to the diametrically opposite corner
of the box. For this test the code was evolved to
a $\delta_{rms}   \approx    1.0$, where $\delta_{rms}$    is  the RMS
variation of the global density  field.  I chose to  run the code only
up  to a      $\delta_{rms}  \approx  1.0$   because  the   Zel'dovich
approximation  breaks down at that  point, making  a comparison to the
approximation impossible.  At the same time spurious velocities in the
two   orthogonal   directions   were    about   $\bar{v}_{y,z}  \simeq
O(10^{-3})$,  which,   when multiplied by  the   total runtime  of the
simulation, corresponds   to spurious   particle   motion in  the  two
orthogonal directions of  less than 1/30th  of a subgrid cell.  Fig. 4
shows a phase   space  plot; the  open  squares are  the  parent  grid
particles, and the solid  dots are the   sub-grid particles. To 
generate this plot I project the particle positions onto the plane which
runs diagonally through the box. I do  not
include the analytic fit because it would obscure most of the sub-grid
points.  Fig. 5 shows the density field for that same simulation. To 
make this plot I simply plotted the density in the cells along the 
diagonal of the box.  This
spurious motion is unavoidable given my use of quasi-isolated boundary
conditions on the sub-grid.  Furthermore  it contributes to the slight
appearance  of  noise   that is     seen in   the phase space    plot.
Nonetheless, the  motion appears  to  be a  small perturbation on  the
overall  collapse of the  pancake, and thus  the effect  appears to be
negligible.

\begin{figure}[t] 
     \epsfxsize = 3.0truein
     \hskip 1.5truein
     \epsfbox{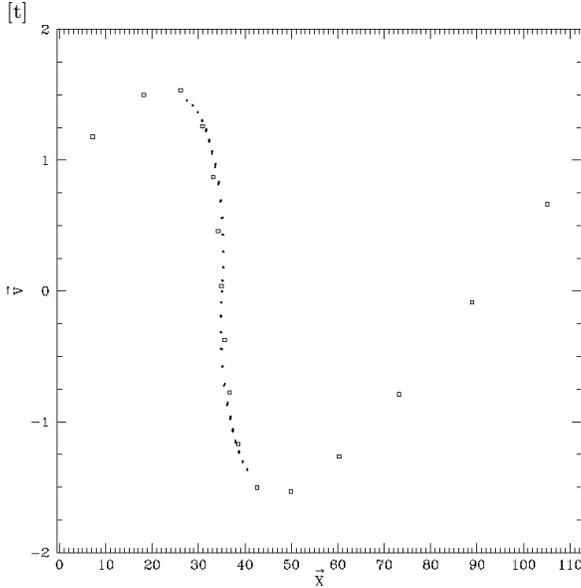}
     \caption{ A phase space plot of the parent grid, and sub-grid particles.
           The open squares are the parent grid particles while the
           open circles are the sub-grid particles. This model was evolved up
           to the onset of non-linearity, $\delta_{rms} = 1.0$.}
\end{figure}

\begin{figure}[t] 
     \epsfxsize = 3.0truein
     \hskip 1.5truein
     \epsfbox{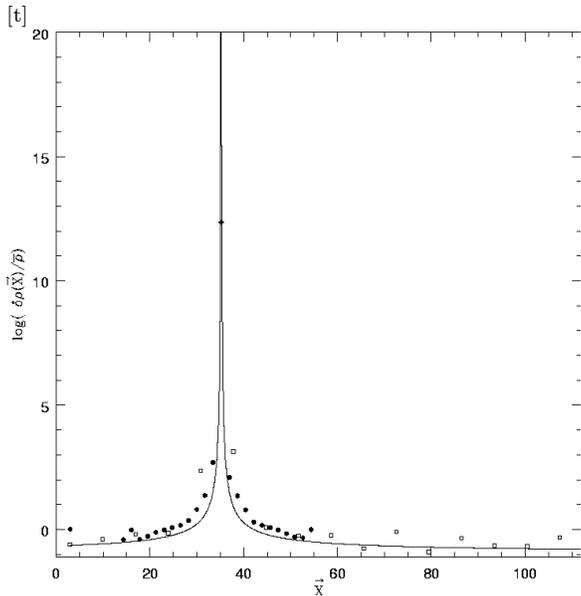}
     \caption{ A plot of the evolved density field on the parent grid and
           the sub-grid at the onset of non-linearity in a 1D collapse
           model. No discernible noise can be detected in the sub-grid
           density due to my method of laying down sub-grid particles.
           The filled in circles are the sub-grid densities, while the 
           open boxes are the parent grid solution.} 
\end{figure}

\subsection{Spherical Infall}

For any   code, comparison with known   analytic results is paramount.
Furthermore, tests which possess a symmetry  not intrinsic to the code
are of particular usefulness. In the spherical infall test I have both
such cases.   The code with its  cubic  cells may tend to  favor tests
which   may possess  planar   symmetry such as   the   plane wave test
above. The   analytic solution to spherical  infall  was worked out in
detail   by  Fillmore    \& Goldreich   1984,   Bertschinger   (1985).
Bertschinger   found for  small values  of   the  radius the  solution
approached a power law  form. In this test  a uniform distribution  of
particles  was laid down   on both the parent  grid  and the sub-grid.
Then I took several time steps in which the density field was given by
the CIC values from the particles plus an extra particle on the parent
grid and 64 extra particles on the sub-grid to serve as seed masses to
start the infall. The  parent seed mass  and the sub-grid seed  masses
all  had the same    coordinates to guarantee    that the infall   was
symmetric with respect to the same point on both grids.  After several
time steps  with this  {\it perturbed}  density field  the extra  seed
masses were removed and the particles were  allowed to move based upon
the CIC  derived  density field  and  the  resulting potential.    The
sub-grid was  chosen to  be 5 parent  grid  cells in size with  $25^3$
sub-grid cells.  This implies  that I can expect  about a factor  of 5
increase in resolution.  The amplitude is fixed uniquely by specifying
the radius at turnaround, Peebles { et  al.} (1989).  Furthermore they
were  able to show given an  initial  mass perturbation $m_0$ then the
turnaround radius, $r_{ta}$ is given by

\begin{equation}
m_0 = { { 9 \pi^3} \over 80} (6 \pi)^{2/3} \rho_b r_{ta}^3.
\end{equation}

\noindent  Therefore once  $r_{ta}$   has  been determined, then   the
dimensionless radius,  $\lambda = r/r_{ta}$,  can be specified and the
normalization  of the density  profile is set.   For this test $r_{ta}
\approx 3.5$ parent grid cells.

To determine the density profile from the particle positions I bin the
particles into spherical shells,   this  gives the number density   of
particles as a function of radius. In the case of the parent particles
this is the density profile $(\rho(r)  - \bar{\rho})/ \bar{\rho}$, but
in the case of the sub-grid particles this  is just the number density
at radius $\lambda$. To  get $(\rho(r) - \bar{\rho})/  \bar{\rho}$ for
the sub-grid  I  must  divide the  number   density by the number   of
sub-grid  particles  per  parent  particle,   this corrects   for  the
$n_{sg}^3$  sub-grid   particles     which    comprise  each    parent
particle. This result gives us the final density distribution.

\begin{figure}[t] 
     \epsfxsize = 3.0truein
     \hskip 1.5truein
     \epsfbox{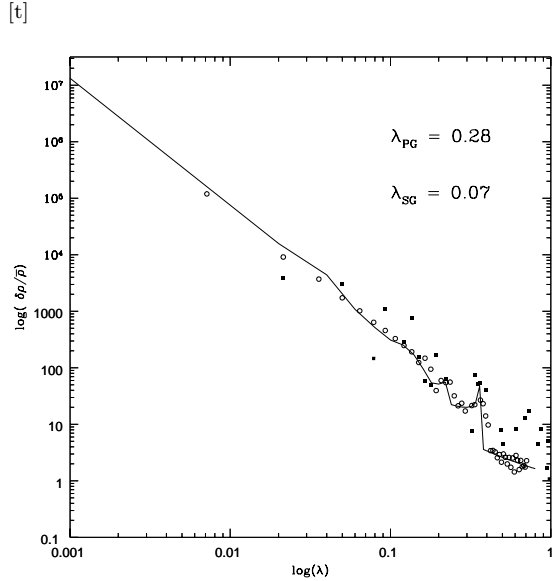}
     \caption{ A comparison of the computed density profiles
              from the nested grid code to the analytic solution. The solid
              line is the analytic solution while the open circles are the 
              sub-grid solution, and the filled in boxes are  the 
              parent grid solution.}
\end{figure}

Fig. 6  is  a plot  of the evolved  density profile  for the spherical
infall problem.  In Fig.  7 and 8 I show   a dot plot of the  particle
positions   within the sub-grid   region. Fig.   7  shows the particle
locations on the  parent grid for a  slice 2 parent cell in thickness,
while  figure  7 shows the  particle  positions on the  sub-grid 1/4 a
parent cell  in  thickness.  In  each  dot plot  one  can  clearly see
spherical  shells. The sub-grid  solution  has some  slight deviations
from perfect    spherical symmetry  near  the   central  peak.   These
deviations could be due to my mass advection scheme, and the fact that
as the particles  enter the sub-grid region  the interpolation used to
assign   sub-grid particle velocities   is    in  error  by a    small
amounts.  Thus  the  particles  are   moving   in slightly  the  wrong
direction.   The effect is  small as  evidence  by the accuracy of the
density profile,   to be considered  in Fig  6. The solid  line is the
Bertschinger analytic  solution,   the open  squares   are the density
profile   of the parent  particles,  and the crosses  are the sub-grid
density profile.   There is some  deviation from the analytic solution
for the parent  particles  which  is  expected since   I  am using   a
relatively coarse grid for this  calculation.  The sub-grid appears to
agree favorably with the analytic  solution down to values of $\lambda
\simeq 0.007$, which corresponds to roughly a length of one-sixth of a
sub-grid cell.  Thus, I appear to  be obtaining better resolution than
one would naively expect from a particle-mesh type code, where I would
expect the sub-grid  solution to be valid  only down to scales  of one
sub-grid cell over   $r_{ta}$, or  0.07.   In  addition,  the sub-grid
solution appears to be roughly  a factor of  10 better than the parent
grid solution. Indicating that  I am getting  better resolution on the
sub-grid by roughly a  factor of 10,  when naively based upon the cell
sizes on the parent grid and sub-grid I  would only expect a factor of
4 in improvement.

\begin{figure}[t] 
     \epsfxsize = 3.0truein
     \hskip 1.5truein
     \epsfbox{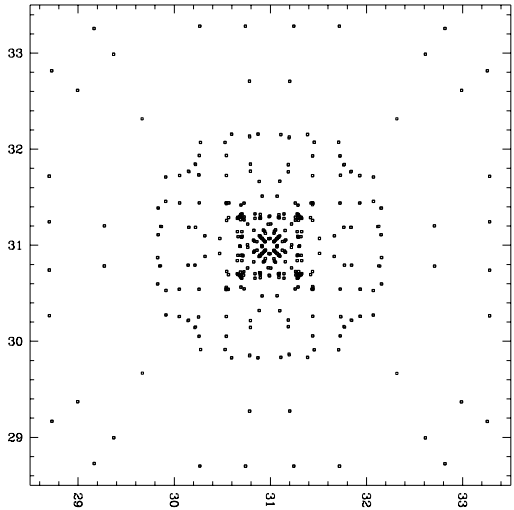}
     \caption{ A slice plot showing the parent particle positions
               on the sub-grid region for a slice two parent cell
               in thickness.}
\end{figure}
\begin{figure}[t] 
     \epsfxsize = 3.0truein
     \hskip 1.5truein
     \epsfbox{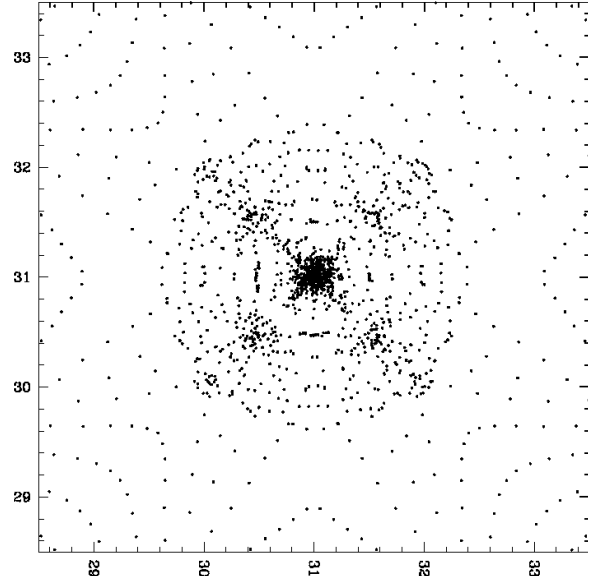}
     \caption{ A slice plot showing the sub-grid particle positions
               on the sub-grid region for a slice 1/4 parent cell 
               in thickness.}
\end{figure}

\subsection{Spherical Hole}

Another  test of how  well the  code  is able  to evolve systems whose
inherent symmetry  differs substantially  from cubic is  the spherical
hole solution  due to  Fillmore \&  Goldreich 1984),  and Bertschinger
(1985). For  this test I run the  code with a single sub-grid centered
on an uncompensated hole.  Bertschinger (1985) showed that the central
void should   grow as  $R \propto  a^n$,   where $n  = 1/5$.     I ran
simulations  using a $16^3$   grid  with the  same  number of   parent
particles  on the grid.  The  parent  particles and sub-grid particles
were then binned  into    spherical  shells to compute   the   density
profiles.  From the  profiles I  find $n  =  0.182 \, \pm  \,  0.022$,
consistent with the theoretical result of Bertschinger.

  

\subsection{Two-Dimensional Perturbations}

For  the final test of   the code I  compare the  results of a $512^2$
two-dimensional  simulation  to the   evolution   of the same  initial
conditions using my  3D nested-grid code. For  this test one of the 2D
models from Beacom, et al. (1991) was used. This  model has an initial
power-law density fluctuation spectra, with a spectral index of 0, and
a high frequency cutoff at $k = 32$. This choice of cutoff was made so
that initially there would not be any power in the Fourier modes which
can only be felt on the sub-grid. Thus, any clustering on the sub-grid
region will be purely a consequence of evolution.

The following test is different from the previous test cases. In all
the previous tests there exists an analytical solution to the test. 
Unfortunately all of the analytic solutions possess certain symmetries.
It is likely that a real model will not possess many of the symmetries
which have been discussed to this point. As a consequence if I wish 
to compare solutions to my nested-grid code to more realistic cases 
I must compare to 2D solutions rather than full 3D. This is due to 
the lack of dynamical range in the 3D simulations. The 2D code used 
here for comparisons has been well tested in the work of Beacom,
et al. (1991), and provides a useful benchmark for more sophisticated 
testing. 

Based  upon  the final evolved  distribution   of particles in the  2D
model, I identify  a sub-grid region.  Then all of the particles which
enter  the  sub-grid  are tagged.   These  particles   will become the
initial  distribution   of sub-grid   particles.   The  full   $512^2$
distribution is  then sampled every fourth  particle to reduce it to a
$64^2$ set of particles. These become the  parent grid particles.  The
$64^2$ distribution of  parent particles are  laid down  onto a $64^3$
grid. I then define a sub-grid region which has the same resolution as
the initial $512^2$ mesh.  The nested-grid code  is then run up to the
same non-linear wavelength, in this case $k_{nl} = 4$, as extrapolated
from linear theory.

I can then  compare  the  distribution  of particles  in  the sub-grid
region  of  the original  2D data  set to  the evolved distribution of
sub-grid particles on the  sub-grid region of  the full 3D simulation.
I should find that the  two distributions agree nicely. Furthermore, I
can use this as a test of collisionality, because as particles accrete
onto one of the clumps in the sub-grid  they should not scatter out of
the plane.   Hence in the final 3D  distribution none of  the sub-grid
particles  should have acquired a  velocity component in the $\hat{z}$
direction.

Fig.   9 is a   $64^2$ sub-sample  of the  full  $512^2$ final evolved
particle distribution for the original 2D perturbations. The left hand
column contains the original 2D particles  while the right hand column
contains   the particles from the nested-grid   3D simulation. The top
left is the evolved  2D distribution plotting  every 8th particle. The
top  right is   a slice  through  the parent  grid   plotting only the
sub-grid region.  The bottom left is the  full set of  2D particles on
the sub-grid  region, while the  lower  right is the  evolved sub-grid
distribution. The bottom two dot pictures of the sub-grid region agree
nicely with only a slight  indication of edge effects. Furthermore  by
comparing the plot in  the upper right to  the plot in the lower right
hand corner one can get a perspective on  the effect of increasing the
mass resolution.

I compare the  initial z positions of the  particles to the  evolved z
positions of the particles by computing the mean of the absolute value
of their difference. I find the motion in the  z direction to of order
1/250{\it th}  of a sub-grid cell,  and thus is completely negligible.
There is no motion on  the parent grid down to  round-off level.  This
indicates that the  quasi-isolated boundary conditions  I apply to the
sub-grid  region   introduce little noise  in  the  form   of sub-grid
particle motion out of the plane of the initial 2D perturbations.

To  make  the comparison   between  the  dot   plots in Fig.    9 more
qualitative I compute  the correlation coefficient between the density
field  on the 2d   sub-grid region and  the  density  field on the  3D
sub-grid.    To generate  a 3D density    field from   the 2D particle
positions I extract all particles on the  sub-grid region plus a small
buffer   to minimize edge effects   and  stack the appropriate  number
vertically  to   make  the    density  field  three-dimensional.   The
correlation coefficient is defined to be

\begin{equation}
K = { {< \delta \rho_{2D}(x) \delta \rho_{SG}(x)>} \over {\sigma_{2D} 
\sigma_{SG}} },
\end{equation}

\noindent where $\delta\rho_{2D}$  is the density field generated from
the 2D distribution, $\delta\rho_{SG}$  is the density field generated
from the sub-grid   distribution of particles, and  $\sigma_{2D}$, and
$\sigma_{SG}$ are  the standard deviations  of the  respective density
fields. For  the model considered  here I find  K = 0.98. Thus  I have
quite good  agreement between  the  original 2D  density field and the
full 3D sub-grid density field.

\subsection{3D Mass Advection Test}

Based upon the 1D Zel'dovich test it is clear that the sub-grid particles
are being advected into the sub-grid region correctly when there is 
little clustering in the sub-grid region. Now I wish to test if the
same conclusion holds when there is clustering of the parent grid 
particles prior to the sub-grid particles being created on the sub-grid.  
There is to my knowledge no analytic solution for such a problem. 
At the request of the referee I am providing this additional test of 
the advection scheme which I am using for the sub-grid particles. 
This test has no known solution and so I will be looking for any 
significant deviations which might be introduced by the advection 
scheme. As I will demonstrate I am unable to find any such effects.  

To begin the test I use a $32^3$ parent mesh with the same number of
particles. The initial spectrum on the parent grid is scale-free and 
has a spectral index of -1. I evolve the system to a non-linear 
wavenumber of 2 performing data dumps at $k_{nl} = 16, 8, 4$,
where $k$ is measured in units of the fundamental. 
Then based upon the $k_{nl}=4$ model I find a region where a merger
has recently taken place between two clumps. I define a sub-grid region
there and evolve the parent grid particles and the sub-grid particles,
again perfomring data dumps at the same non-linear wavenumbers. 
In figure 10 I have plotted a series of dot plots through the 
sub-grid region. The right-hand column is the sub-grid grid particles
and the left-hand plots are the parent particles. The top row are
the particle positions at $k_{nl} = 16$, the middle row at $k_{nl}=8$,
and the bottom row is at $k_{nl}=4$. In all cases the sub-grid 
distributions appear to be consistent with the parent grid particles.

Once again to make the arguement more qualitative I compute the
density cross-correlation function between the sub-grid particles and
the parent particles in several ways. First I extract all the
parent and sub-grid particles within the sub-grid region plus a 
small buffer to avoid edge effects. I bin both sets of particles 
onto a mesh the same resolution as the parent mesh. I then 
cross-correlate these density fields. These results are in column
2 of table 1, and are labeled $K_{SG}$. The third column of table 
1 contains the cross-correlations
cimputed in a slightly different manner. Since I know from which 
parent particle each sub-grid particle came, I can compute the center of
mass of the cloud of sub-grid particles for each parent particle on 
the sub-grid. I then bin these centers of mass onto a mesh with the 
same resolution as the parent mesh. I then compute the cross-correlation
between this new density field obtained from the sub-grid particle
centers of mass and the parent grid particles as before. These 
cross-correlations are in column 3 of table 1, and are labeled $K_{CM}$.

There is relatively good agreement in all cases. Of course I don't
expect the cross-correlation to be exact. The sub-grid has much 
higher resolution than the parent grid so the sub-grid particles 
will have small scale motions which will cause deviations in their
motions from what one would expect without higher force resolution. 
Furthermore, the agreement between the sub-grid distribution and the
parent grid distribution gets poorer as the models evolve further. 
This is to be expected due the presence of small scale motions on
the sub-grid which cannot be resolved on the parent grid. In addition,
the second set of cross-correlations is better than the first which 
again is to be expected since the center of mass of a cloud of 
sub-grid particles should remain reasonably close to the location of
the corresponding parent particle. Any deviations will be due to the
small scale motions which can exist on the sub-grid and not the 
parent grid. All of this is consistent  with the results of Melott 
\& Shandarin (1993), who  find that if one computes the cross-correlation
between models whose only difference is the cutoff scale  $k_{c}$ 
then in the limit that $k_{nl} << k_{c}$ the difference between
the two models should become negligible.

\section{Acknowledgments}

I would like to thank NASA for support through a NASA Graduate Student
Fellowship,  NGT50816, the  National Science Foundation (AST-9021414),
my thesis advisor Adrian L.  Melott for many valuable discussions, and
Jennifer Pauls, Capp Yess, and Sergei Shandarin for a critical reading
of the manuscript prior to submission. All calculations were performed
at  the National  Center  for  Supercomputing Applications  in Urbana,
Illinois.

\clearpage
\onecolumn

\begin{figure}
     \hskip 1.5truein
     \epsfbox{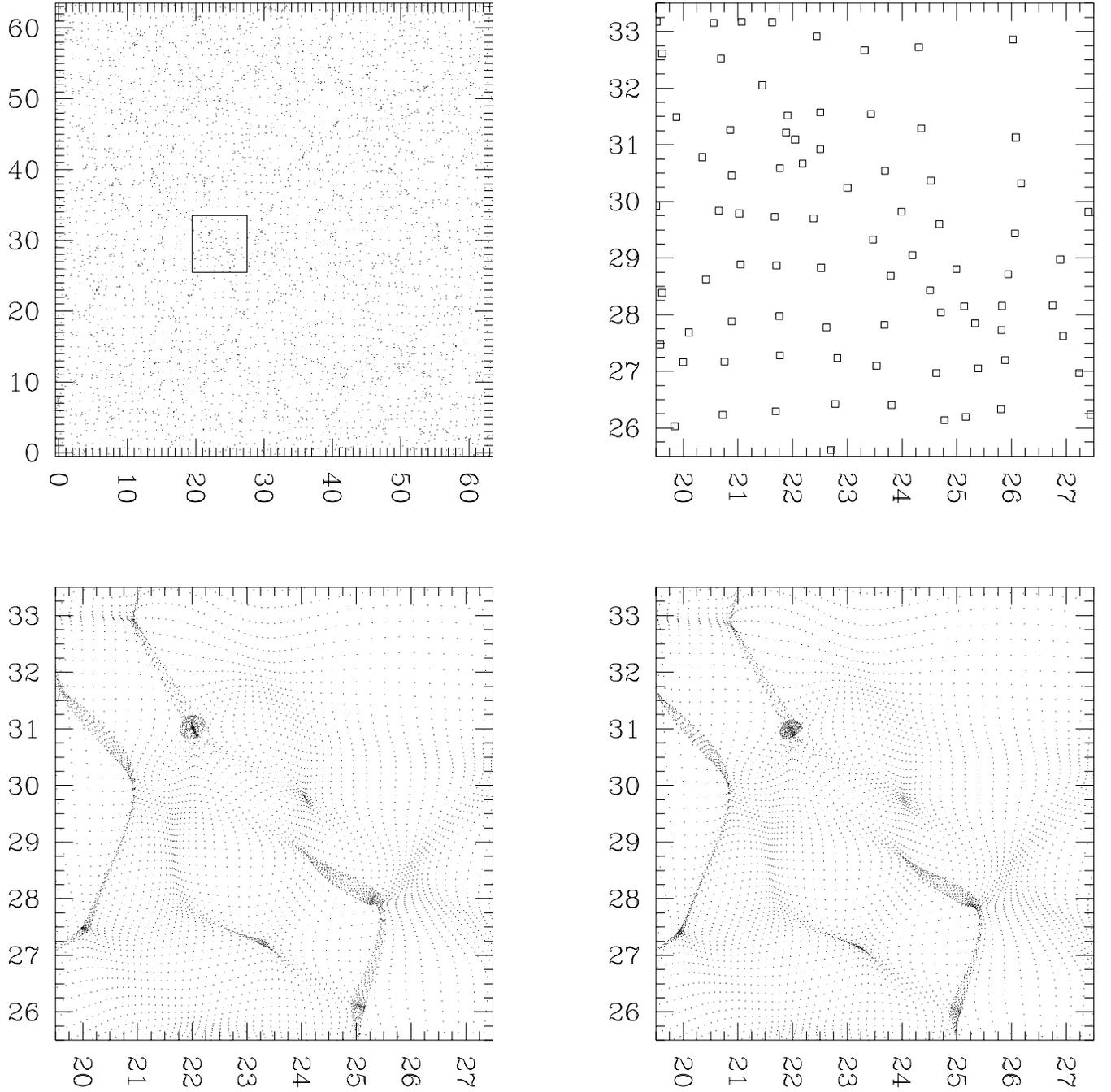}
     \caption{A dot plot of a slice through the distribution of 
              particles in the 2D perturbations test. The left hand 
              column contains the original
              2D distribution, and the right hand column contains
              the 3D distribution from the nested-grid code. The upper 
              left plot is the evolved distribution of particles in the
              2D case, where only a $64^2$ subset of the original $512^2$
              particles has been plotted. The small box labels the position
              of the sub-grid region with the parent grid. The upper right 
              is a slice 
              through the 
           sub-grid region of the parent grid particles. 
              The lower left is the evolved distribution of particles
              from the 2D distribution on the sub-grid region. The lower
              right is the evolved distribution of sub-grid particles.
              The small boxes in the upper left plot labels the sub-grid 
              region.}
\end{figure}

\begin{figure}
     \hskip 1.5truein
     \epsfbox{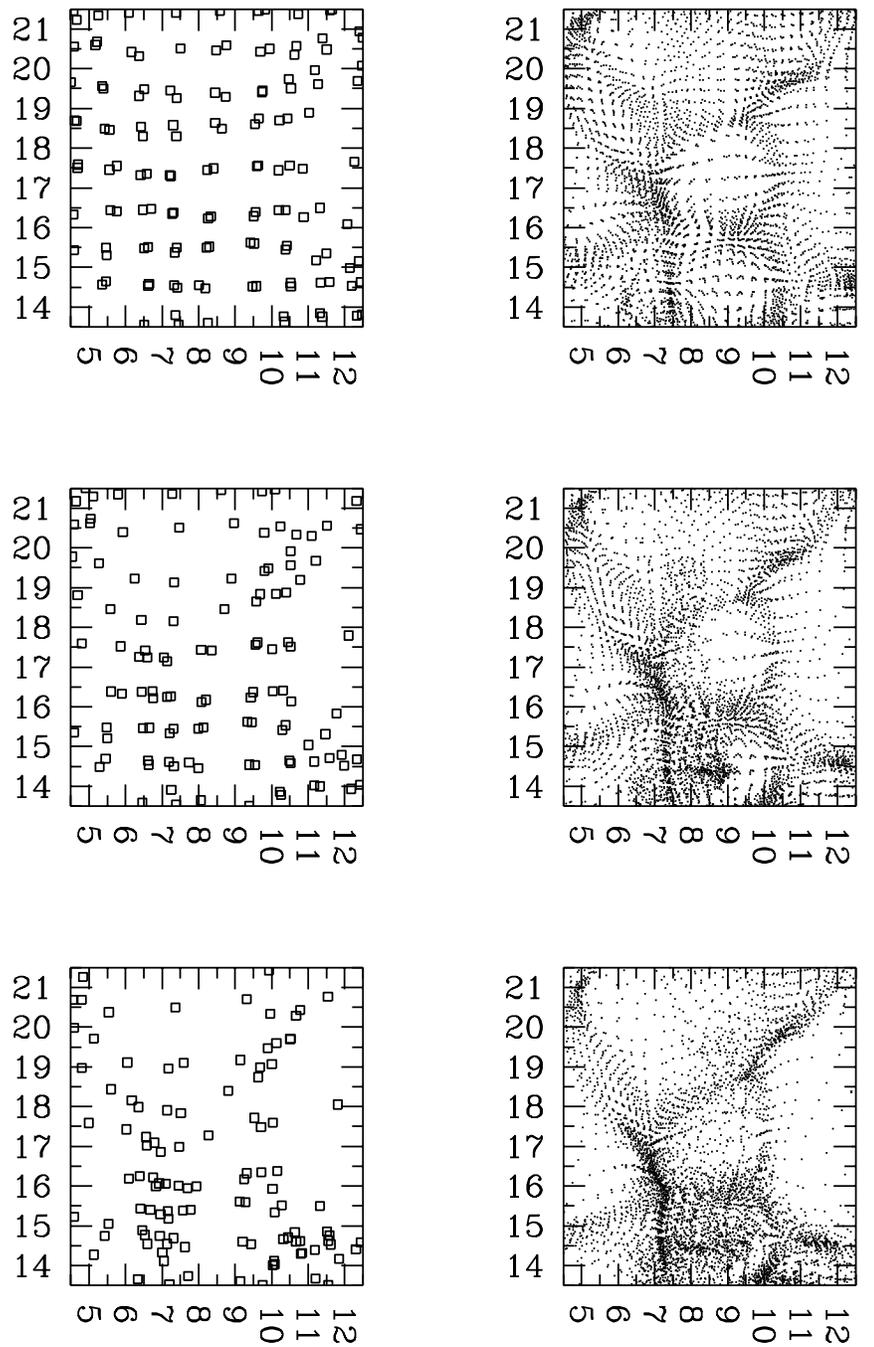}
     \caption{The left-hand column contains plots of the parent 
              particles in a 
              slice 2 parent grid cells thick. The right hand 
              column is the sub-grid particle positions in a 
              slice 1 parent cell in thickness. The top row
              are the particle positions at $k_{nl} = 16$, the
              middle row are the particles at $k_{nl} = 8$,
              and the bottom row are the particles at $k_{nl} =4$.} 
\end{figure}

\onecolumn
\newpage

%

\onecolumn
\newpage

\begin{table}
\caption{Cross-Correlations from the 3D Mass Advection Test} 

\begin{center}
\begin{tabular}{ccl} \hline \hline
$k_{nl}$ &  $K_{SG}$ & $K_{CM}$ \\
\cline{1-3}
16 &  0.942 & 0.971    \\
8  &  0.893 & 0.942   \\
4  &  0.874 & 0.897   \\
\end{tabular}
\end{center}
\end{table}

\end{document}